\documentclass[twocolumn,showpacs,preprintnumbers,amsmath,amssymb]{revtex4}
\usepackage{graphicx}
\usepackage{textcomp}

\begin{document}
\draft
\title{Bloch oscillations in complex crystals with $\mathcal{PT}$ symmetry}
  \normalsize

\author{S. Longhi}
\address{Dipartimento di Fisica and Istituto di Fotonica e
Nanotecnologie del CNR, Politecnico di Milano, Piazza L. da Vinci 32, I-20133 Milano, Italy}


%
\bigskip
\begin{abstract}
\noindent Bloch oscillations (BO) in complex lattices
 with $\mathcal{PT}$ symmetry are theoretically investigated with specific reference to optical BO
 in photonic lattices with gain/loss regions. Novel dynamical phenomena with no counterpart in ordinary
lattices, such as non-reciprocal BO related to violation of the
Friedel's law of Bragg scattering in complex potentials, are
highlighted.
\end{abstract}



\maketitle

Bloch oscillations (BO), i.e. the coherent oscillatory motion of a
quantum particle in a periodic potential driven by an external dc
force, represent one of the most striking predictions of wave
mechanics of periodic systems. In the quantum realm, BO have been
experimentally observed for electrons in semiconductor superlattices
\cite{Leo98} and for ultracold atoms or Bose-Einstein condensates
 in tilted optical lattices \cite{BEC}. Classical analogues of BO
have been also proposed and observed for optical
\cite{OBOwaveguides} and acoustical \cite{acoustical} waves. The
phenomenon of BO is related to the transition of energy spectrum
from continuous (energy bands) to (nearly) discrete, with the
formation of Wannier Stark (WS) ladders when the dc force is
applied. In physical space, BO are generally explained in terms of
wave Bragg scattering off the periodic potential, causing a wave
packet to oscillate rather than translate through the lattice.
Previous studies on BO have considered particle motion in
real-valued potentials, may be in presence of nonlinearity, lattice
disorder or particle interactions \cite{Bishop96}. However, particle
or wave dynamics in complex potentials is not uncommon in physics
\cite{Hatano96,Keller97,Siegman97,Berry04}. Complex potentials for
matter waves emerge in the near resonant interaction of light with
open two-level systems \cite{Keller97}, whereas in optics wave
propagation in lossy or active waveguides is described by a complex
refractive index \cite{Siegman97}. Complex potentials have received
in recent years an enormous interest since the discovery that
non-Hermitian Hamiltonians with parity-time ($\mathcal{PT}$)
symmetry may possess an entirely real-valued energy spectrum below a
phase transition (symmetry-breaking) point \cite{Bender}.
$\mathcal{PT}$ Hamiltonians belong to the more general class of
pseudo-Hermitian systems \cite{Mostafazadeh02}. Optical realizations
of $\mathcal{PT}$-symmetric potentials, based on light propagation
in coupled waveguides with a complex refractive index, have been
recently proposed \cite{Ganainy07,Makris08,Klaiman08}, whereas
energy band formation and spontaneous symmetry-breaking thresholds
have been investigated in a few $\mathcal{PT}$ potential models
\cite{bande} and in presence of nonlinearity \cite{Musslimani08} or
lattice disorder \cite{Shapiro09}. It is the aim of this Letter to
investigate the onset of BO in periodic lattices with $\mathcal{PT}$
symmetry, highlighting some unique features arising from the complex
nature of the scattering potential. We consider here BO for light
waves in photonic lattices \cite{OBOwaveguides}, however the
analysis can be applied to other physical systems, such as to matter
waves in accelerating complex optical lattices.
\par
Paraxial propagation of light at wavelength $\lambda$ in a
weakling-guiding one-dimensional waveguide lattice with a
superimposed transverse refractive index gradient is described by
the Schr\"{o}dinger-type equation \cite{Makris08,Longhi09}
\begin{equation}
i \lambdabar \partial_z \psi= -\frac{\lambdabar^2}{2n_s}
\partial^2_x \psi+V(x)\psi +Fx \psi \equiv (\mathcal{H}_0+Fx)\psi
\end{equation}
where $\lambdabar=\lambda/(2 \pi)$ is the reduced wavelength, $n_s$
is the substrate refractive index, $V(x)=n_s-n(x)$ is the potential,
$n(x)=n(x+a)$ is the refractive index profile of the lattice
(spatial period $a$), and $F$ is the refractive index gradient which
mimics an external dc force. For $F=0$, $\mathcal{PT}$ symmetry
requires
 $V(-x)=V^*(x)$, i.e. suitable combinations of optical gain and loss regions in the
lattice, as discussed in \cite{Makris08,Klaiman08,Ganainy07}. The
real and imaginary parts of the potential are denoted by $V_R(x)$
and $\alpha V_I(x)$, respectively, where $\alpha \geq 0$ is a
dimensionless parameter that measures the anti-Hermitian strength of
$\mathcal{H}_0$. For given profiles of $V_R$ and $V_I$, the spectrum
of $\mathcal{H}_0$ turns out to be real-valued and composed by bands
separated by gaps for $\alpha < \alpha_{c}$, where $\alpha_c \geq 0$
corresponds to the spontaneous symmetry breaking point. For $\alpha
\geq \alpha_c$, band merging and appearance of pairs of
complex-conjugates eigenvalues is observed \cite{Makris08,bande}.
For instance, for the potential
\begin{equation}
V_R(x)=V_0 \cos(2 \pi x/a), \; V_I(x)=V_0 \sin(2 \pi x/a)
\end{equation}
one has $\alpha_c=1$ \cite{Makris08}. \par Let us first consider the
case $\alpha<\alpha_c$, and indicate by
$\phi_n(x,\kappa)=u_n(x,\kappa) \exp(i \kappa x)$ the Bloch-Floquet
eigenfunction of $\mathcal{H}_0$, corresponding to the $n$-th
lattice band and to wave number $\kappa$ in the first Brillouin zone
($-\pi/a
 \leq \kappa < \pi/a$) with periodic part $u_n(x,\kappa)$. Thus $\mathcal{H}_0
 \phi_n(x,\kappa)=E_n(\kappa)\phi(x,\kappa)$, where $E_n(\kappa)$ is the
 real-valued band dispersion curve.
  As for real potentials, one can show that
 $E_n(-\kappa)=E_n(\kappa)$. Additionally, $\mathrm{Re}(u_n(\kappa,x))$ and $\mathrm{Im}(u_n(\kappa,x))$
 have well defined and opposite parity under the inversion $x \rightarrow -x$,
 whereas $\phi^{\dag}_n(x,\kappa)=\phi_n(-x,-\kappa)$ are eigenfunctions of the adjoint $\mathcal{H}_{0}^{\dag}$ with same
 eigenvalues
 $E_n(\kappa)$. Assuming that $ I_n \equiv \int_{0}^{a} dx u^{*}_n(-x,-\kappa) u_n(x,\kappa)$ does not vanish, i.e. in absence of spectral singularities,
 the orthonormal conditions
 $\langle \phi^{\dag}_l(x,\kappa')| \phi_n(x,\kappa) \rangle \equiv \int_{-\infty}^{\infty}dx \phi^{\dag *}_l(x,
 \kappa') \phi_n(x,\kappa)= \mathcal{D}_n \delta_{n,l} \delta(\kappa-\kappa') $ hold, where $\mathcal{D}_n=\pm 1$ is given by the sign of $I_n$
in each band  \cite{Bender,Makris08}. To investigate WS
localization, like in ordinary crystals we expand the wave packet
$\psi$ as a superposition of Bloch-Floquet states \cite{Callaway},
$\psi(x,z)=\sum_n \int d \kappa c_n(\kappa,z) \phi_n(\kappa,z)$. The
scalar product of both sides of Eq.(1) with $\langle
\phi_{n}^{\dag}(x,\kappa)|$ yields the following evolution equations
for the spectral coefficients $c_n(\kappa,z)$
\begin{equation}
i \lambdabar \left( \partial_z-\frac{F}{\lambdabar}
\partial_{\kappa} \right)c_n = E_n(\kappa)c_n+F \mathcal{D}_n \sum_l X_{n,l}(\kappa)
c_l
\end{equation}
where $X_{n,l}(\kappa) \equiv (2 \pi i /a) \int_{0}^{a}dx
u_{n}^{*}(-x,-\kappa) \partial_{\kappa} u_l(x,\kappa)$. The
off-diagonal elements $X_{n,l}$ ($ n \neq l$) in Eq.(3) are
responsible for interband transitions, i.e. Zener tunneling (ZT),
and can be written in the equivalent form $X_{n,l}=(2 \pi
\lambdabar^2/ a n_s) [E_n(\kappa)-E_l(\kappa)]^{-2} \int_0^a
u_n^*(-x,-\kappa)
\partial_x V(x) u_l(x,\kappa)$.
If bands $n$ and $l$ are separate by a large gap and the force $F$
is small enough such that $F |X_{n,l}(\kappa)| \ll
|E_{n}(\kappa)-E_l(\kappa)|$ in the entire Brillouin zone , ZT is
negligible as in ordinary lattices \cite{Callaway}. Taking $X_{n,l}
\simeq 0$ for $n \neq l$ (single-band approximation), the WS
spectrum of the $n$-th band is then obtained by letting in Eq.(3)
$c_n(\kappa,z)=a_n(\kappa) \exp(-iE z/ \lambdabar)$ and imposing the
boundary condition $a_n(2 \pi/a)=a_n(0)$. This yields
\begin{equation}
 E_l^{(n)}=Fla+\langle E_n\rangle +i \gamma_n \; \; \; (l=0, \pm
1, \pm 2,...) \; ,
\end{equation}
 where we have set $\langle E_n\rangle = (a/ 2 \pi) \int_{- \pi/a}^{ \pi/a} d\kappa
E_n(\kappa)$ and $\gamma_n= (\mathcal{D}_n F a/ 2 \pi) \int_{
-\pi/a}^{\pi/a} d\kappa \mathrm{Im} (X_{n,n}(\kappa))$. As compared
to an ordinary crystal \cite{Callaway}, the main difference here is
that the WS ladder spectrum is complex-valued, even though $\alpha <
\alpha_c$. This is simply due to the fact that the external force
$F$ breaks the $\mathcal{PT}$ symmetry of the Hamiltonian. In each
ladder, the imaginary parts of the eigenvalues have the same value
(i.e., independent of $l$), equal to $\gamma_n$. Since
$\mathrm{Im}(X_{n,n}(\kappa))$ is an even function of $\kappa$,
$\gamma_n$ is usually non-vanishing, and its sign is reversed as the
direction of the force, i.e. the sign of $F$, changes. Therefore, in
complex lattices below the phase transition point, BO  with spatial
period $z_B=\lambda/(Fa)$ are observed, however they are either
amplified or damped depending on the sign of $F$. This behavior is
illustrated in Fig.1, which shows the onset of amplified or damped
BO as obtained by numerical simulations of Eq.(1) for the complex
lattice of Eq.(2) with $\alpha=0.3$ and for an input excitation of
the lattice with a broad Gaussian beam at normal incidence. In the
figures, the evolution of the total beam power $P=P(z)$ is also
shown. Note that, in spite all WS eigenvalues have the same
imaginary part, $P(z)$ shows a non-exponential behavior, with abrupt
changes of $P(z)$ in correspondence of Bragg scattering of the beam
from the lattice. This behavior is ascribable
to the non-orthogonality of WS states in complex potentials.\\
\begin{figure}[htbp]
  \includegraphics[width=85mm]{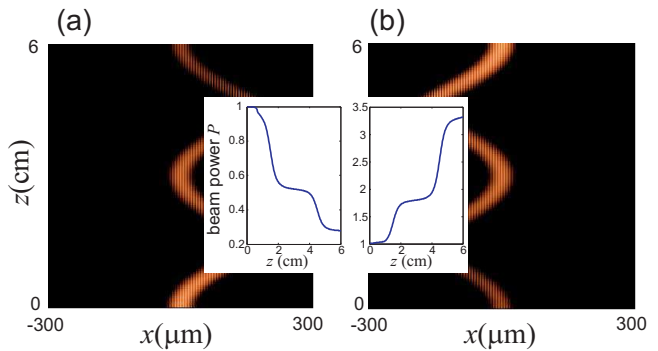}\\
   \caption{(color online) BO in the complex lattice  of Eq.(2) for $\alpha=0.3$ and for an index gradient
   (a) $F=-3.52 \; \mathrm{m}^{-1}$, and (b) $F=3.52 \; \mathrm{m}^{-1}$.
   Other parameter values are $a=6 \; \mu\mathrm{m}$, $\lambda=633 \; \mathrm{nm}$, $V_0=0.001$, and $n_s=1.42$. The figures show
   the evolution of field amplitude $|\psi(x,z)|$ and beam power $P=\int dx |\psi(x,z)|^2$ (normalized to the input power)
   versus propagation distance $z$. The BO period is $z_B=3$ cm.}
\end{figure}
 As the non-Hermitian
parameter $\alpha$ is increased to cross the phase transition value
$\alpha_c$, gap narrowing till band merging, associated to the
appearance of pairs of complex-conjugate eigenvalues, is observed
\cite{Makris08}, making the previous analysis inapplicable. Here we
will limit to investigate the onset of BO in the $\alpha = \alpha_c$
and $\alpha>\alpha_c$ regimes by considering two specific examples
which enable an analytical
treatment.\\
\begin{figure}[htbp]
  \includegraphics[width=85mm]{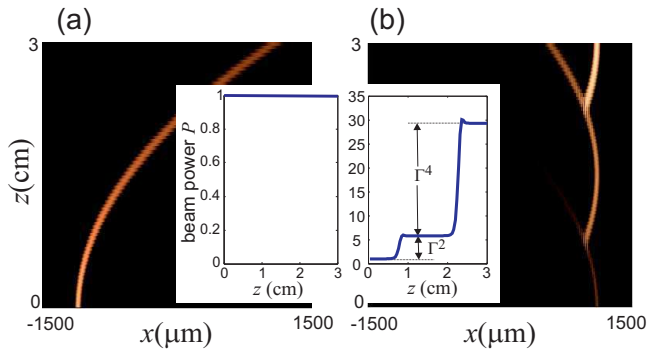}\\
   \caption{(color online) Same as Fig.1, but for $\alpha=\alpha_c=1$ and
   (a) $F=-7.03 \; \mathrm{m}^{-1}$, (b) $F=7.03 \; \mathrm{m}^{-1}$. Other parameter values are as in Fig.1, except for
   $V_0=2 \times 10^{-4}$. Note that BO disappear for a negative index gradient $F$.}
\end{figure}
 The first example is given by the potential (2) at the
phase transition point $\alpha=\alpha_c=1$, i.e. $V(x)=V_0 \exp(2
\pi i x/a)$ \cite{Berry98}, for which band gaps disappear and the
energy dispersion is described by the free-particle parabola
$E(\kappa)=(\lambdabar^2/2n_s) \kappa^2$, periodically folded inside
the first Brillouin zone. In presence of an external force $F$,
numerical simulations indicate that the wave packet evolution
strongly depends on the sign of $F$, as shown in Fig.2. Remarkably,
BO disappear for $F<0$, and the wave packet propagates following a
parabolic path as if the complex lattice were absent [Fig.2(a)].
This result indicates that a net drift along the lattice can be
achieved by the application of a dc force, an effect which is not
possible in ordinary crystals without collisions. Conversely, for
$F>0$ an oscillatory dynamics survives, with spatial period equal to
$z_B=\lambda/(Fa)$ [Fig.2(b)]. In this case, at propagation
distances corresponding to Bragg reflection from the lattice, a
rather abrupt amplification of the scattered beam is observed [see
the inset of Fig.2(b)]. This dynamical regime, which can be referred
to as unidirectional BO, has no counterpart in ordinary lattices and
is related to violation of Friedel's law of Bragg scattering by
complex potentials \cite{Keller97,Berry98}, i.e. breakdown of Bragg
diffraction invariance by crystal inversion. Analytical insights
into unidirectional BO can be gained by writing Eq.(1) in momentum
space and assuming that the lattice is excited by a broad wave
packet at normal incidence, i.e. $\psi(x,0)=\int dk g(k)
\exp(-ikx)$, where the plane-wave spectrum $g(k)$ is narrow at
around $k=0$ with a spectral width much smaller than the Bragg wave
number $k_B=2 \pi/a$. Let us indicate by $\psi_0(x,z)$ the
propagated wave packet that one would observe in absence of the
lattice, i.e. the solution to Eq.(1) with $V=0$. The center of mass
$\langle x \rangle (z)$ of $\psi_0(x,z)$ follows the parabolic path
$\langle x \rangle (z)=-(F/2n_s)z^2$ due to the external force $F$,
whereas the mean value of momentum $\langle k \rangle$ grows
linearly, $\langle k \rangle (z)=-Fz$. In presence of the complex
lattice, Bragg diffraction sets in when $\langle \kappa \rangle$
gets close to $-k_B/2$, but not when $\langle \kappa \rangle \sim
k_B/2$; this nonreciprocal behavior of Bragg scattering by complex
potentials is discussed e.g. in \cite{Keller97,Berry98}. In presence
of the lattice, one can show that the wave packet is expressed by a
superposition of retarded wave packets $\psi_0(x,z-nz_B)$ with
amplitudes $B_n(z)$, namely
\begin{equation}
\psi(x,z) \simeq \sum_{n=0}^{+\infty}B_n(z) \psi_0(x,z-nz_B).
\end{equation}
In Eq.(5), $B_0(z)=1$, whereas for $n \geq 1 $ the amplitudes $B_n$
can be iteratively calculated from the relation
\begin{equation}
 B_n(z)=-i\frac{V_0}
{\lambdabar} \int_0^z d \xi \exp[ -i \varphi_n(\xi)] B_{n-1}(\xi),
\end{equation}
where we have set $\varphi_n(z)=(\lambdabar^2 k_B/2 n_sF)
[(Fz/\lambdabar-k_Bn)^2+k_B(Fz/\lambdabar-k_Bn)+k_B^2/3]$ . In the
limit $z_B \gg (n_s a/ \pi F)^{1/2}$ and for $F<0$, $|B_{n \geq
1}(z)|$ remains small because of $\varphi_n(z)$ has no stationary
points for $z>0$. Physically, this corresponds to the absence of
Bragg scattering (the condition $\langle k \rangle \sim -k_B/2$ is
never satisfied), and the scenario of Fig.2(a) is observed.
Conversely, for $F>0$ the phase $\varphi_n$ has a stationary point
at $z=z_B(n-1/2)$, and one obtains $|B_n(z)| \sim \Gamma^n H(z-n
z_B+z_B/2)$, where $\Gamma \equiv (V_0/ \lambdabar) (n_s a/F)^{1/2}$
and $H(\xi)$ is the Heaviside step function [$H(\xi)=0$ for $\xi<0$,
$H(\xi)=1$ for $\xi>0$]. In this case, Eq.(5) clearly shows that a
periodic regeneration and amplification of the wave packet $\psi_0$
due to Bragg scattering occurs at $z=z_B/2, 3 z_B/2, 5z_B/2,...$,
explaining the behavior observed in Fig.2(b). For parameter values
used in the simulations, the predicted value of amplification factor
$\Gamma$ is $\Gamma \simeq 2.186$, with which the staircase behavior
of beam power shown in the inset of Fig.3(b) is reproduced with
excellent accuracy.
\par As a second example, we consider a complex binary lattice \cite{Musslimani08} with
 broken $\mathcal{PT}$ symmetry  at $\alpha_c=0$, and show that the external force can restore a real-valued energy
 spectrum. The lattice consists of an array of
equally-spaced single-mode waveguides with alternating gain
($V_I>0$) and loss ($V_I<0$) channels, i.e. $V_R(x)=\sum_l
h(x-la-a/2)$ and $V_I(x)=\sum_l (-1)^l h(x-la-a/2)$, where
$h(x)=h(-x)$ defines the (real) index profile of the waveguide
channel (see Fig.3(a)). Note that, for $\alpha \neq 0$, the lattice
period is $2a$. A typical band diagram is shown in Fig.3(b). In the
tight-binding approximation and for  $\alpha \ll 1$, light transport
in the lattice  is described by coupled-mode equations for the modal
amplitudes  $a_l$ of light trapped in the waveguides
\cite{OBOwaveguides,Longhi09,Musslimani08,Dreisow09}
\begin{equation}
i \frac{da_l}{dz}=-\frac{\Delta}{4} (a_{l+1}+a_l)+f l a_l+i(-1)^l
\frac{g}{2} a_l
\end{equation}
\begin{figure}[htbp]
  \includegraphics[width=85mm]{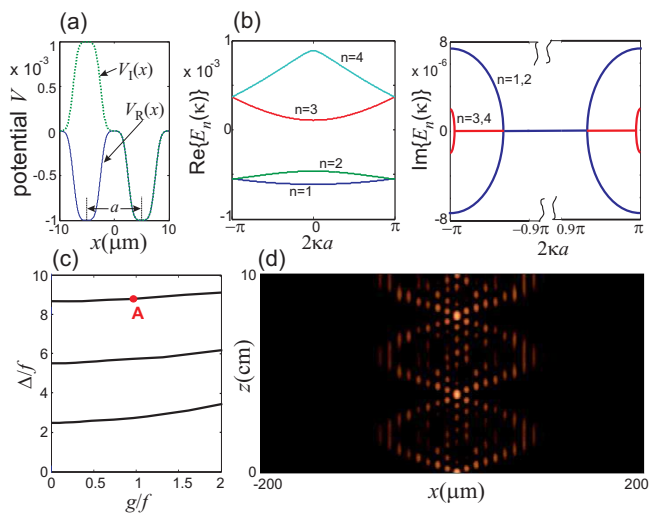}\\
   \caption{(color online) BO in a complex binary lattice ($a=10 \; \mu$m). (a) Profiles of $V_R$ and $V_I$, and
  (b) corresponding band diagram for $\alpha=0.01$ (real and imaginary parts of energy $E$). (c) Resonance curves in the
   $(\Delta/f,g/f)$ plane corresponding to a real-valued WS ladder
   spectrum. (d) Breathing BO mode (single waveguide input excitation) for
   $F=1.618 \; \mathrm{m}^{-1}$, corresponding to a Bloch period $z_B \simeq 39.1 \; \mathrm{mm}$.}
\end{figure}
where $\Delta/4$ is the real-valued coupling rate between adjacent
waveguides, $f=Fa/ \lambdabar$, and $g$ is the real-valued power
gain/loss coefficient in the waveguides, which is proportional to
$\alpha$. For the potential of Fig.3(a) with $\alpha=0.01$, from the
band diagram of Fig.3(b) (the two merged bands $n=1$ and $n=2$) one
can estimate $g \simeq 1.46 \; \mathrm{cm}^{-1}$ and $\Delta \simeq
14 \; \mathrm{cm}^{-1}$. For $F=0$, the spectrum of Eq.(7) is purely
continuous and composed by two merged minibands [the two lowest
bands of Fig.3(b)], namely \cite{Musslimani08}
\begin{equation}
E(\kappa)= \pm \lambdabar \sqrt{(\Delta/2)^2 \cos^2( \kappa
a)-(g/2)^2} \; \;, |\kappa| \leq \pi/ 2 a.
\end{equation}
Since complex-conjugate eigenvalues appear near the band edges for
any value of $g>0$, the lattice has zero threshold for
$\mathcal{PT}$ symmetry-breaking, i.e. $\alpha_c=0$. In presence of
the external force $F$, the spectrum becomes purely discrete and
composed by two interleaved WS ladders $E_{l}^{(1)}$ and
$E_{l}^{(2)}$, as in a binary lattice with real potential
\cite{Dreisow09}. Specifically, one can show that
\begin{equation}
E_l^{(1,2)}=2Fal-\frac{F a \varphi_{1,2}}{\pi} + i\frac{Fa \;
\mathrm{ln} |y_{1,2}|}{\pi}
\end{equation}
($l=0, \pm1, \pm 2, ...$) where $y_{1,2} \equiv |y_{1,2}| \exp(i
\varphi_{1,2})=\mathrm{Re}(\sigma) \pm
[1+\mathrm{Re}(\sigma)^2]^{1/2}$, $\sigma \equiv \mathcal{S}_{1,2}$,
and the $2 \times 2$ matrix $\mathcal{S}$ is the propagator for the
differential system $dx_{1,2}/d \kappa= \pm i (\Delta /4f)
\cos(\kappa /2)x_{1,2}+(g/4f) x_{2,1}$ from $\kappa=0$ to $\kappa=2
\pi$. Note that the two WS ladders are generally complex-valued with
opposite sign of their imaginary part. However, if the condition
$\mathrm{Re(\sigma)}=0$ is satisfied, $y_{1,2}= \pm1$ and the
 spectrum becomes real-valued and given by a single WS ladder,
namely $E_l=Fal$ ($l=0, \pm 1, \pm 2,...$). Figure 3(c) depicts the
numerically-computed locus in the $(\Delta/f,g/f)$ plane where
$\mathrm{Re(\sigma)}=0$, showing the existence of a sequence of
resonance curves. A perturbative calculation of the propagator
$\mathcal{S}$ in the $g \rightarrow 0$ limit shows that the
resonance curves depart from the points of the $ \Delta/f$ axis
 that are the roots of Bessel function $J_0$ \cite{Dreisow09}.
Along such resonance curves, a forced Bloch particle undergoes
undamped nor amplified BO, with spatial period $z_B=\lambda/(Fa)$.
This is shown in Fig.3(d), which depicts an example of a BO
breathing mode, obtained by numerical analysis of Eq.(1) for single
waveguide excitation of the complex binary lattice of Fig.3(a), with
a forcing $F=1.618 \; \mathrm{m}^{-1}$, corresponding to point A
($g/f \simeq 0.91$, $\Delta/f \simeq 8.72$) on the third resonance
curve of Fig.3(c).\\
Photonic $\mathcal{PT}$ lattices may provide an accessible
laboratory to experimentally observe BO in complex crystals. Complex
refractive indices could be realized in engineered
 arrays of active semiconductor waveguides with selective pumping
to achieve gain and loss regions. At optical or near-infrared
wavelengths, a modulation depth of the imaginary refractive index
$\Delta n_I \sim 2 \times 10^{-4}$  (as in Fig.2) requires a
gain/absorption coefficient of the order $\sim$ 10-20 cm$^{-1}$,
which is available with current semiconductor optical amplifier
technology. Smaller gain/loss coefficients are needed for the
example of Fig.3. Non-reciprocal BO could be also observed in
passive (i.e. without gain) waveguide arrays with loss modulation,
in a geometrical setting similar to the one recently realized to
demonstrate $\mathcal{PT}$ symmetry breaking in a passive AlGaAs
optical coupler \cite{Salamo09}. The transverse index gradient can
be realized by either chirping the waveguide width
\cite{OBOwaveguides} or by waveguide axis bending \cite{Longhi09}.\\
 In conclusion, BO in complex crystals with
$\mathcal{PT}$ symmetry behave differently than in ordinary lattices
owing to non-reciprocity of Bragg scattering. Our results indicate
that the conventional wisdom of BO and transport for classical or
matter waves needs to be modified when dealing with complex
crystals.


\begin{thebibliography}{31}


\bibitem{Leo98}
C. Waschke, H. G. Roskos, R. Schwedler, K. Leo, H. Kurz, and K.
K\"{o}hler, Phys. Rev. Lett. {\bf 70}, 3319 (1993).


\bibitem{BEC}
M. BenDahan, E. Peik, J. Reichel, Y. Castin, and C. Salomon, Phys.
Rev. Lett. {\bf 76}, 4508 (1996); S. R. Wilkinson, C. F. Bharucha,
K. W. Madison, Q. Niu, and M. G. Raizen, Phys. Rev. Lett. {\bf 76},
4512 (1996); B.P. Anderson and M. A. Kasevich, Science {\bf 282},
1686 (1998).

\bibitem{OBOwaveguides}
R. Morandotti, U. Peschel, J. S. Aitchison, H. S. Eisenberg, and Y.
Silberberg, Phys. Rev. Lett. {\bf 83}, 4756 (1999); T. Pertsch, P.
Dannberg, W. Elflein, A. Br\"{a}uer, and F. Lederer, Phys. Rev.
Lett. {\bf 83}, 4752 (1999); H. Trompeter, T. Pertsch, F. Lederer,
D. Michaelis, U. Streppel, A. Br\"{a}uer, and U. Peschel, Phys. Rev.
Lett. {\bf 96}, 023901 (2006).

\bibitem{acoustical}
H. Sanchis-Alepuz, Y.A. Kosevich, and J. Sanchez-Dehesa, Phys. Rev.
Lett. {\bf 98}, 134301 (2007).
%

\bibitem{Bishop96}
D. Cai, A.R. Bishop, N. Gr{\o}nbech-Jensen, and M. Salerno, Phys.
Rev. Lett. {\bf 74}, 1186 (1995); F. Dominguez-Adame, V.A. Malyshev,
F.A.B.F. de Moura, and M.L. Lyra, Phys. Rev. Lett. {\bf 91}, 197402
(2003); A. Buchleitner and A.R. Kolovsky, Phys. Rev. Lett. {\bf 91},
253002 (2003).

\bibitem{Hatano96}
N. Hatano and D.R. Nelson, Phys. Rev. Lett. {\bf 77}, 570 (1996).

\bibitem{Keller97}
C. Keller, M.K. Oberthaler, R. Abfalterer, S. Bernet, J.
Schmiedmayer, and A. Zeilinger, Phys. Rev. Lett. {\bf 79}, 3327
(1997).

\bibitem{Siegman97}
A. Kostenbauder, Y. Sun, and A.E. Siegman, J. Opt. Soc. Am. A {\bf
14}, 1780 (1997).

\bibitem{Berry04}
 M.V. Berry, Czech. J. Phys. {\bf 54}, 1039 (2004).

\bibitem{Bender}
C.M. Bender and S. Boettcher, Phys. Rev. Lett. {\bf 80}, 5243
(1998); C.M. Bender, Rep. Prog. Phys. {\bf 70}, 947 (2007).

\bibitem{Mostafazadeh02}
A. Mostafazadeh, J. Math. Phys. {\bf 43}, 2814 (2002).

\bibitem{Ganainy07}
R. El-Ganainy, K. G. Makris, D. N. Christodoulides, and Z. H.
Musslimani, Opt. Lett. {\bf 32}, 2632 (2007).

\bibitem{Makris08}
K.G. Makris, R. El-Ganainy, D.N. Christodoulides, and Z.H.
Musslimani, Phys. Rev. Lett. {\bf 100}, 103904 (2008).

\bibitem{Klaiman08}
S. Klaiman, U. G\"{u}nther, and N. Moiseyev, Phys. Rev. Lett. {\bf
101}, 080402 (2008).

\bibitem{bande}
C.M. Bender, G.V. Dunne, P.N. Meisinger, Phys. Lett. A {\bf 252},
272 (1999); H.F. Jones, Phys. Lett. A {\bf 262}, 242 (1999); Z.
Ahmed, Phys. Lett. A {\bf 286}, 231 (2001); A. Khare and U.
Sukhatme, J. Math. Phys. {\bf 47}, 062103 (2006); T. Curtright and
L. Mezincescu, J. Math. Phys. {\bf 48}, 092106 (2007).

\bibitem{Musslimani08}
Z.H. Musslimani, K.G. Makris, R.El-Ganainy, and D.N.
Christodoulides, J. Phys. A {\bf 41}, 244019 (2008).

\bibitem{Shapiro09}
O. Bendix, R. Fleischmann, T. Kottos, and B. Shapiro, Phys. Rev.
Lett. {\bf 103}, 030402 (2009).

\bibitem{Longhi09}
S. Longhi, Laser \& Photon. Rev. {\bf 3}, 243 (2009).

\bibitem{Callaway}
J. Callaway, {\it Quantum Theory of the Solid State} (Academic
Press, New York, 1974), pp. 465-478; V. Grecchi and A. Sacchetti,
arXiv:quant-ph/0506057.

\bibitem{Berry98}
M.V. Berry, J. Phys. A {\bf 31}, 3493 (1998).

\bibitem{Dreisow09}
B.M. Breid, D.Witthaut, and H.J. Korsch, New J. Phys. {\bf 8}, 110
(2006); F. Dreisow, A. Szameit, M. Heinrich, T. Pertsch, S. Nolte,
A. T\"{u}nnermann, and S. Longhi, Phys. Rev. Lett. {\bf 102}, 076802
(2009).

\bibitem{Salamo09}
A. Guo, G.J. Salamo, D. Duchesne, R. Morandotti, M. Volatier-Ravat,
V. Aimez, G. A. Siviloglou, and D. N. Christodoulides, Phys. Rev.
Lett. {\bf 103}, 093902 (2009).

\end{thebibliography}

\end{document}